%% file: 2016banfi_sdn-multipath.tex
\documentclass[letter,conference,10pt]{IEEEtran}

\usepackage[utf8]{inputenc}
\usepackage[T1]{fontenc}
\usepackage[australian]{babel}
\usepackage{color,xcolor}
\usepackage{listings}
\usepackage{subfig}
\usepackage{graphicx}
\usepackage{url}
\usepackage{multirow}
\usepackage{graphicx}
\usepackage{amssymb}
\usepackage{amsfonts}
\usepackage{booktabs}
\usepackage{xspace}
\usepackage{verbatim}
\usepackage{footnote}
\usepackage{epstopdf}
\usepackage{algorithm}
\usepackage{algorithmic}

\newcommand{\theTitle}{Endpoint-transparent Multipath Transport with Software-defined Networks}
\newcommand{\theAuthor}{Dario Banfi, Olivier Mehani, Guillaume Jourjon, Lukas Schwaighofer, Ralph Holz}
\newcommand{\theKeywords}{multipath transport, SDN, OpenFlow, Open vSwitch}

\newif\ifhypercrap
\hypercraptrue
\ifhypercrap
    \usepackage[unicode]{hyperref}
    \hypersetup{pdfstartview=FitV,
      colorlinks=true, linkcolor=black, citecolor=black, urlcolor=blue,
      pdftitle={\theTitle}, pdfauthor={\theAuthor}, pdfkeywords={\theKeywords}
}
\fi

\usepackage[maxnames=3,style=ieee,firstinits=true,doi=false,sortcites,isbn=false,backend=bibtex]{biblatex}
\usepackage{booktabs}
\usepackage[detect-weight=true, binary-units=true, group-digits=true, group-separator={,}]{siunitx}

\newif\ifstatus
\statustrue

\input{macros.tex}

\title{\theTitle}
\author{
  \IEEEauthorblockN{
    Dario Banfi\IEEEauthorrefmark{1}\IEEEauthorrefmark{3},
    Olivier Mehani\IEEEauthorrefmark{1},
    Guillaume Jourjon\IEEEauthorrefmark{2},
    Lukas Schwaighofer\IEEEauthorrefmark{3},
    Ralph Holz\IEEEauthorrefmark{4}
  }\\
  \IEEEauthorblockA{
  \IEEEauthorrefmark{1} NICTA, Sydney, Australia; Email: olivier@mehani.name \\
  \IEEEauthorrefmark{2} Data61, CSIRO, Sydney, Australia; Email: \{first.last\}@data61.csiro.au \\
    \IEEEauthorrefmark{3} Faculty of Informatics, Technical University of Munich, Germany; Email: \{first.last\}@tum.de \\
    \IEEEauthorrefmark{4} School of IT, University of Sydney, NSW, Australia; Email: ralph.holz@sydney.edu.au
  }
}

\begin{document}

\maketitle

\begin{abstract}

Multipath forwarding consists of using multiple paths simultaneously to
transport data over the network. While most such techniques require
endpoint modifications, we investigate how multipath forwarding can be done
inside the network, transparently to endpoint hosts. With such a
network-centric approach, packet reordering becomes a critical issue as it
may cause critical performance degradation.

We present a Software Defined Network architecture which automatically sets
up multipath forwarding, including solutions for reordering and performance
improvement, both at the sending side through multipath scheduling
algorithms, and the receiver side, by resequencing out-of-order packets in
a dedicated in-network buffer.

We implemented a prototype with commonly available technology and evaluated
it in both emulated and real networks. Our results show consistent
throughput improvements, thanks to the use of aggregated path capacity.  We
give comparisons to Multipath TCP, where we show our approach can achieve a
similar performance while offering the advantage of endpoint transparency.

\end{abstract}


\input{docs/introduction.tex}

\input{docs/related.tex}

\input{docs/architecture.tex}

\input{docs/evaluation.tex}

\input{docs/discussion.tex}

\input{docs/conclusion.tex}

\printbibliography

\end{document}

%% file: macros.tex
\newcommand{\SIms}[1]{\SI{#1}{\milli\second}}
\newcommand{\SIgbps}[1]{\SI[per-mode = symbol, per-symbol = /]{#1}{\giga\bit\per\second}}

\newcommand{\SImbps}[1]{\SI[per-mode = symbol, per-symbol = /]{#1}{\mega\bit\per\second}}

\newcommand{\SIMB}[1]{\SI{#1}{\mega\byte}}

\newcommand{\latinlocution}[1]{\textit{#1}}

\newcommand{\eg}{\latinlocution{e.g.}\xspace}
\newcommand{\ie}{\latinlocution{i.e.}\xspace}

%% file: docs/introduction.tex
\section{Introduction}
\label{sec: introduction}

IP networks are inherently multipath. Yet, the existence of multiple paths
between two endpoints is rarely leveraged. This issue can be ascribed to the
fact that only lower layers can establish an accurate view of the network
topology, while only upper layers are able to control transmission rate and
end-to-end connectivity.

Nonetheless, solutions have been proposed at various layers to enable specific
use-cases and improve performance. Examples are given at layers 2--3 for
data-centres with, \eg, BCube~\cite{guo2009bcube} or
DCell~\cite{zhang2010cloud}, or at layer 4 for multi-homed devices with
Multipath TCP (MPTCP)~\cite{rfc6824} or Concurrent Multipath Transfer for SCTP
(CMT-SCTP)~\cite{2010yuan_CMT-SCTP_parallel_subflows}.

The most prominently quoted motivations for multipath are the potential for
continuity of connectivity in case of path failure or congestion (\ie,
fail-over or load-balancing), or capacity aggregation to speed up high volume
transfers between endpoints~\cite[\eg,][for MPTCP]{rfc6182}.

Layer-2 multipath topologies~\cite[\eg,][]{al2008scalable}, have been
successfully deployed and used within fully-controlled data-centre networks.
End-to-end multipath support throughout the public Internet is however
limited~\cite{2015mehani_mptcpscan} due to the requirement to modify end-hosts.
Heterogeneous network paths also worsen the issue of packet reordering,
creating head-of-line blocking delays, and sometimes leading to worse
performance than single-path
transfers~\cite{2012sarwar_performance_multipath_assymetric}.

In this paper,\footnote{This paper improves on the first author's MSc thesis
but focuses on TCP only; please refer
to~\cite{2016banfi_endpoint-transparent_multipath_sdn_msc} for more details and
other transport protocols.} we attempt to join both lower- and upper-layer
approaches and merge their successes through the use of SDN. We aim to satisfy
the following goals: capacity aggregation, ease of end-to-end deployment,
adaptivity to failures, and automatic path computation. To this end, we
introduce the MPSDN architecture, comprising an SDN controller with better
knowledge and control of available paths than endpoint-only layer-4 solutions,
as well as modifications of the Open vSwitch implementation and OpenFlow
protocol to enable finer packet scheduling and reordering within the network,
without need for explicit end-host support.

The solution can be deployed with either layer-2 forwarding or layer-3 routing
or tunnelling, and the controller does not require full control of the network
hops. We show that this approach enables performance similar to MPTCP's
while lifting the requirement for end-host modifications. The focus of this
paper is on TCP, but we note that our proposal can handle other transport
protocol in a similar
fashion~\cite{2016banfi_endpoint-transparent_multipath_sdn_msc}. Our work also
allows us to identify some non-trivial issues when implementing layer-4
switching and scheduling with SDN solutions.

The proposed mechanism can offer benefits in several scenarios where additional
bandwidth would enhance the Quality of Experience for users.  A typical
scenario is high-definition video streaming where the bit-rate is higher than
the capacity of a single path.\footnote{A video demonstration of this use-case
can be found at \url{https://www.youtube.com/watch?v=hkgf7l9Lshw}} Another
use-case for this proposal is that of multi-cloud overlay networks between
virtualised environments.\footnote{See, for example, Docker's overlays
\url{https://docs.docker.com/engine/userguide/networking/get-started-overlay/}.}
In this scenario, a user controls the edges of the network and deploys the
proposed mechanism to maximise bandwidth utilisation between clouds.

The remainder of this paper is organised as follows: The next section reviews
state of the art of multipath approaches in line with the goals of our
research. We present the proposed architecture and its implementation in
Section~\ref{sec:architecture} and provide a performance evaluation in both
emulated conditions and in a real multi-homed testbed in
Section~\ref{sec:evaluation}. We give some insight and lessons learnt about
mixing SDN and multipath in Section~\ref{sec:discussion} and offer concluding
remarks in Section~\ref{sec:conclusion}.

%% file: docs/related.tex
\section{Related Work}
\label{sec:related}

Multipath topologies in both layer 2 and layer 3 networks are common, offering
multiple communication options for capacity aggregation, load-balancing, and
congestion avoidance. This section reviews the state-of-the-art of solutions
proposed to leverage those capabilities. We do this layer by layer, from 2 to
4, and offer some insight about previous uses of SDN for this purpose.

Table \ref{tab:relatedwork} summarizes the discussed work in light of our design
goals. With ``easy deployability'' we denote the use of software/hardware that
can be incrementally deployed and used on real networks and that is not only an
experimental proof-of-concept.

\begin{table}[bt]
  \addtolength{\tabcolsep}{-1.5pt}
  \centering
  \caption{Comparison of characteristics and fulfilment of our goals of
  state-of-the-art multipath proposals and MPSDN.}
  \label{tab:relatedwork}
  \begin{tabular}{lccccccccc}
    \toprule
    & \rotatebox{90}{\textbf{TRILL}}      & \rotatebox{90}{\textbf{SPB}}    &
    \rotatebox{90}{\textbf{FLARE}}      & \rotatebox{90}{\textbf{Harp}}     &
    \rotatebox{90}{\textbf{MPTCP}}      & \rotatebox{90}{\textbf{CMT-SCTP}} &
    \rotatebox{90}{\textbf{AMR}}        & \rotatebox{90}{\textbf{OLiMPS}}   &
    \rotatebox{90}{\textbf{MPSDN}} \\
    \midrule
    Layer 2                    & \checkmark & \checkmark & \checkmark &		   & \dag	&            & \checkmark & \checkmark & \checkmark \\
    Layer 3                    &            &            & \checkmark & \checkmark &            &            &            &            & \checkmark \\
    Layer 4                    &            &            &            &		   & \checkmark & \checkmark &            &            & \ddag \\
    SDN-based solution	       &            &            &            &            &            &            & \checkmark & \checkmark & \checkmark \\
    \midrule
    Bandwidth Aggregation      &            &            &            & \checkmark & \checkmark & \checkmark & \checkmark & \checkmark & \checkmark \\
    Easy Deployability         & \checkmark & \checkmark &            &            &            &            & \checkmark & \checkmark & \checkmark \\
    Adaptivity to Failures     & \checkmark & \checkmark &            & \checkmark &            & \checkmark & \checkmark &            & \checkmark \\
    Load-Balancing             & \checkmark & \checkmark & \checkmark & \checkmark &            & \checkmark & \checkmark & \checkmark & \checkmark \\
    Automatic Path Computation & \checkmark & \checkmark &            &            &            &            & \checkmark & \checkmark & \checkmark \\
    \bottomrule
  \end{tabular}
  \dag\  MPTCP was used for aggregation on top of a multihomed L2 network\\
  \ddag\ MPSDN uses L4 knowledge, \eg, sequence numbers, to reorder packets
\end{table}

\subsection{Link-layer Multipath}

The spanning tree (STP) protocol is extensively used on L2 networks to ensure
loop-free forwarding in Ethernet networks. It has the downside of actively
pruning paths from the networks which could be utilized for increased
bandwidth.  Cisco's layer-2 multipath~\cite{l2mp} attempts to remediate this by
enabling the use of alternate paths, while the IEEE 802.3ad amendment
introduces provisions for link aggregation~\cite{iee802.3}. Neither solution
however offers full multipath support across complex topologies.

TRILL (Transparent Interconnection of Lots of Links)~\cite{rfc5556} uses IS-IS
routing to ensure that every bridge has full knowledge of the network, allowing
for the creating of an optimal forwarding tree with support for Equal-Cost
Multipath (ECMP)~\cite{rfc2991}. 802.1aq SPB (Shortest Path
Bridging)~\cite{spb} also leverages IS-IS to compute a shortest path through
the network. A designated MAC address (used with SPB-MAC) or VLAN ID (SPB-VID)
is assigned for each switch, and used as label on  each received frames.
Packets travel on the shortest path to the edge switch, which again
de-encapsulates the frame and sends it to the end device.  Neither of these
techniques allows aggregated bandwidth because of their use of ECMP-like
hashing.

MPTCP, discussed in more details below, has also been suggested as a way to
leverage multiple layer-2 paths in data-centres and improve performance and
robustness~\cite{RBPGWH11}. It has been shown that, with a sufficiently high
number of subflows, it is possible to aggregate capacity and increase
load-sharing. The downsides of this approach are the necessary end-host
support, the lack of multipath capability for other protocols such as UDP or
SCTP, and its limitation to data-centres.

\subsection{Network-layer multipath}

Flowlet Aware Routing Engine (FLARE)~\cite{Kandula2007} is a dynamic multipath
load balancing technique.  It uses time delays between packets of the same flow
to split them into \emph{flowlets} that may be distributed on different paths.
This allows to distribute the traffic between available paths more
accurately, as compared to flow-based distribution, while maintaining in-order
arrival at the receiver.

FLARE has shown, through trace-driven simulations of tier-1 and regional ISPs,
that highly accurate traffic splitting can be implemented with very low state
overhead and negligible impact on packet reordering. However its focus is on
load-balancing and does not offer capacity aggregation.

The Harp network architecture prioritizes foreground traffic and uses multipath
to dissipate background transfers~\cite{4215742}.

It can leverage path diversity and load imbalance in the Internet to tailor
network resource allocation to human needs (foreground vs.\ background
traffic).  It also provides better fairness and utilization compared to
single-path end-host protocols. Moreover, it can be deployed at either
end-hosts or enterprise gateways, thereby aligning the incentive for deployment
with the goals of network customers.  Packet reordering is performed at the
exit gateways to cope with different path latencies. Its focus on background
traffic at the exception of all other traffic, however, makes it ill-fitted for
our goals.

\subsection{Transport-layer multipath}

Extensions to two main transport protocols have been proposed to support
multipath. MPTCP~\cite{rfc6824} introduced a new set of TCP options to enable
negotiation between multipath-capable hosts while using backward-compatible TCP
packets on each path. SCTP's fail-over supports
load-balancing~\cite{2004yi_model_load_balancing_sctp} and has been extended to
support concurrent multipath transfer~\cite{2006iyengar_cmt-sctp}. Despite
their intrinsic limitation to a single transport protocol, those approaches
have seen reasonable success in the lab, with their main barrier to deployment
being the need for end-host support.

A very active area of research with transport-layer multipath is enabling
packet schedulers to deal with path asymmetry without introducing head-of-line
blocking~\cite{2012sarwar_performance_multipath_assymetric}. Most schedulers
attempt to distribute packets unevenly or out-of-order across available paths,
so they arrive in order at the
destination~\cite{2012singh_comparison_scheduling_multipath,2013sarwar_daps,2014kuhn_daps,2014paasch_experimental_evaluation_MPTCP_schedulers,2014yang_out-of-order_transmission_in-order_arrival_scheduling_MPTCP,2015le_forward_delay_scheduling_mptcp,2016ferlin_blest_blocking_estimation_mptcp_scheduler}.
An adequate scheduling policy is important to enable the benefits of capacity
aggregation in heterogeneous scenarios.

\subsection{SDN-based multipath solutions}

Multipath in OpenFlow has been proposed back in
2010\footnote{\url{http://archive.openflow.org/wk/index.php/Multipath_Proposal}}
and later implemented through \emph{Groups} (such as Select or All) to enable
L2/L3 multipath forwarding for load-balancing purposes.  It has since then been
researched extensively~\cite{li2013openflow,Koerner2014118,6407519}, but none
of these approaches allows for aggregated bandwidth as they all rely on flow
hashing (as does OpenFlow at its core).

Adaptive Multipath Routing (AMR) has been used to perform layer-2 aggregation
in data-centres~\cite{2015subedi_openflow-based_multipath_aggregation}. It
splits flows over multiple paths and introduces an architecture which adapts
dynamically to network congestion and link failures.  An interesting aspect of
this approach is its computation of max-flow paths throughout the network to
determine the best combination to use. An analogous technique has also been
used in~\emph{OLiMPS} (OpenFlow Link-layer Multipath Switching) to utilize
robust inter-domain connectivity over multiple physical links.

\vspace{1em}

Overall, existing proposals can either not provide aggregated path capacity or
are limited to layer-2 forwarding. Layer-4 solutions, while supporting
aggregation as their main advantage, lack in deployability as they require
end-host support. AMR comes closest to our goals, but is an L2-only solution.
Moving forwards, we propose an architecture able to handle both layer-2 and -3
multipath scenarios, while accounting for the scheduling and reordering
requirements of heterogeneous paths.

%% file: docs/architecture.tex
\section{Architecture}
\label{sec:architecture}

\begin{figure}
\centering
\includegraphics[width=\linewidth]{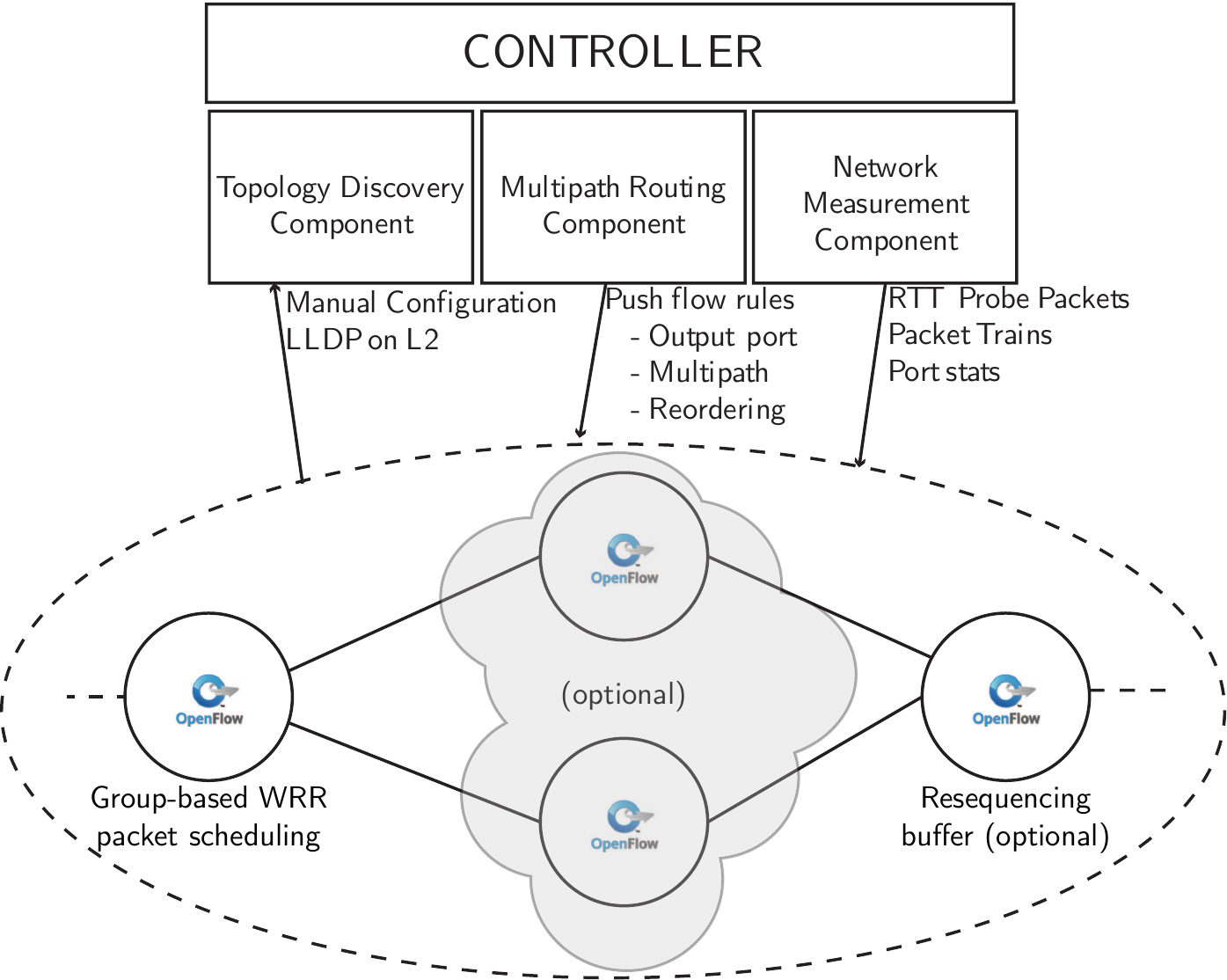}
\caption{Multipath SDN Architecture}
\label{fig:mpsdn_architecture}
\end{figure}

Our proposed architecture for an endpoint-transparent multipath network
consists of a centralized controller with knowledge of the network topology
which dynamically sets up loop-less forwarding rules on SDN switches under its
control (\figurename~\ref{fig:mpsdn_architecture}). For the presented
proof-of-concept, we focus on two path scenarios only.

The controller has some knowledge of the network state and views the underlying
infrastructure as a directed graph, where costs between switches are given by
the latency and capacity of the paths.  With this knowledge it computes the
optimal multipath forwarding table to send data from one node to the other,
maximizing the capacity usage with an algorithm based on the maximum-flow
problem. This is similar to
AMR~\cite{2015subedi_openflow-based_multipath_aggregation}, but we extend it to
layer-3 infrastructures.  In case of failure or heavy congestion, the
controller will compute an updated forwarding table and push it to the SDN
switches.

In the remainder of this section, we present the key concepts of our
architecture: the topology discovery and path selection, as well as the packet
scheduler and reordering buffer.  We also describe how we implemented this
architecture in the Ryu OpenFlow
controller\footnote{\url{https://osrg.github.io/ryu/}} and how we modified Open
vSwitch to support packet reordering on edge switches.

\subsection{Topology Discovery}

In order to discover the network topology, we both query the forwarding devices
using the Link Layer Discovery Protocol (LLDP) when available (\ie, layer 2) or
deploy ad hoc mechanisms to estimate end-to-end latency and throughput (\ie,
layer 3). In particular, we estimate path latency with a slightly modified
Bouet's algorithm~\cite{Bouet2013}, which yields high accuracy and has a low
network footprint. Unlike NetFlow or measurements using ICMP echo requests,
this does not require additional servers or components.  The algorithm is run
using controller-to-switch messages only.

We use port statistics counters for bandwidth estimation. As shown in
OpenNetMon~\cite{VanAdrichem2014}, we can accurately monitor a flow's
throughput by probing flow statistics periodically.  The controller uses a
similar approach by periodically requesting port statistic messages from its
switches (every 2 seconds in the current implementation). The per-port
available capacity is determined by subtracting the maximum capacity with the
utilization from the last period of observation.

\subsection{Path Selection}

In order to maximize the aggregated capacity of multiple paths, the controller
uses an algorithm similar to the Edmonds-Karp algorithm to solve the
\emph{maximum flow} problem, with a Breadth First Search to find the augmenting
paths.  It uses the Dijkstra algorithm with min-priority queue to find the
shortest paths from source to destination.  The estimated available bandwidth
between the nodes is used to maximize the overall throughput between the sender
and the receiver.

Pilot experiments showed that, in a similar manner as for layer-4 multipath,
not all paths are compatible and a very high delay imbalance was detrimental.
To select compatible paths, we introduce the concept of \emph{maximum delay
imbalance},
\begin{equation}
    \mathrm{MDI} = \frac{d_{max}}{d_{max}+d_{min}}-0.5\,,
  \label{eq:max_delay_imbalance}
\end{equation}
where $d_{min}$ and $d_{max}$ denote the minimal and maximal delays from the
candidate paths, and $0.5$ a rescaling factor.
Its range is $[0, 0.5]$, where $0$ represents completely balanced paths and $0.5$ is
the limit of imbalance.

This metric is used for different purposes in our solution.  If the computed
$\mathrm{MDI}$ among the selected paths is higher than a reordering threshold,
a flow reordering rule is set up at the receiving edge router. Similarly, if
the $\mathrm{MDI}$ is above another threshold, the delay difference is
considered too high to provide any aggregated capacity advantage. We determine
those thresholds in Section~\ref{sec:evaluation}

\subsection{Packet Scheduler}

The common challenge for every multipath protocol is deciding how to send data
over the available paths.  The task is usually done by a \emph{scheduling
algorithm}.  This scheduler can rarely work in isolation as it needs to adapt
to changing path characteristics, mainly in terms of delays and congestion.
There are many approaches to multipath
scheduling~\cite{2012singh_comparison_scheduling_multipath}, ranging from
simple information agnostic round-robin approaches to omniscient algorithms.

To maximize the performance, a multipath scheduler should push the right amount
of data over different paths, without overloading already congested ones and
ensuring full utilization of the available capacity. MPTCP uses subflows with
independent congestion windows~\cite{rfc6182,rfc6824} and can buffer some
packets before sending them on the desired
path~\cite{2014yang_out-of-order_transmission_in-order_arrival_scheduling_MPTCP,
2016ferlin_blest_blocking_estimation_mptcp_scheduler}.

In the case of in-network multipath, however, neither the per-path window
information nor the advance buffering option are readily available.  To
maximize application throughput, we use a Weighted Round-Robin (WRR) scheduler
which sends bursts of packets along the paths, weighted according to their
capacity as $w_j = c_j/\sum_{i}^{n}{c_i}$, where $w_j$ is the weight associated
with path \emph{j}, and $c_j$ its estimated capacity.
While not as fine-grained as layer-4 scheduling, this approach maps well to
OpenFlow's Groups approach and our measurements, presented in
Section~\ref{sec:evaluation}, show the performance difference is acceptable.

\subsection{Reordering Mechanism}

By selecting multiple paths with potentially different characteristics, our
mechanism introduces packet reordering.  To avoid a performance impact due to
out-of-order packets, we implemented a corrective mechanism that can be
deployed on the edge switches.

Layer-4 multipath algorithms (Section~\ref{sec:related}) solve this problem by
using out-of-order queues at the receiver, which resequence packets in the
desired order prior to passing them to the application.

We introduce a \emph{resequencing buffer} at the receiving edge switch in order
to address this problem in a similar fashion, albeit without the receiver
node's involvement. The buffer temporarily stores packets received ahead of
time.  It does so by maintaining a record of the next expected sequence number
for each flow, in a similar fashion as TCP, and only forwards packets if the
sequence numbers match. This is show in Algorithm~\ref{alg:reordering_algo}.

\begin{algorithm}
\caption{Resequencing for each flow}
\label{alg:reordering_algo}
\begin{algorithmic}
  \REQUIRE buffer $B$ of size $S$
  \REQUIRE buffering threshold $T$
  \REQUIRE loss-recovery factor $LRF$
  \WHILE{$pkt \leftarrow$ receive packet}
  \IF{$pkt$ is SYN}
  \STATE $expected \leftarrow  pkt.seq + pkt.size$
  \STATE forward $pkt$
  \ELSIF{$pkt.seq < expected$}
  \STATE forward $pkt$ \COMMENT{immediately forward duplicates}
  \ELSIF{$pkt.seq = expected$}
  \FORALL{$p \in B | p.seq < expected$}
  \STATE forward $p$ \COMMENT{send all delayed packets in order}
  \ENDFOR
  \STATE forward $pkt$
  \STATE $expected \leftarrow expected + pkt.size$
  \ELSIF {$B.use > T$}
  \STATE store $pkt$ in $B$
  \FORALL{$p \in B$}
  \STATE forward $p$ \COMMENT{send all packets in order, ignoring gaps}
  \STATE $lastp <- p$
  \ENDFOR
  \STATE $expected \leftarrow lastp.seq + lastp.size \cdot LRF$ \COMMENT{account for bursty losses}
  \ELSIF {$B.use < S$}
  \STATE store $pkt$ in $B$
  \ELSE
  \STATE $spkt \leftarrow p \in B | p.seq = \min_{p\in B}(p.seq)$
  \IF{$spkt.seq \le pkt.seq$}
  \STATE send $spkt$ \COMMENT{send the packet with the lowest sequence number}
  \STATE store $pkt$ in $B$
  \ELSE
  \STATE forward $pkt$
  \ENDIF
  \ENDIF
  \ENDWHILE
\end{algorithmic}
\end{algorithm}

This can cause a problem in case packets are lost prior to reaching the
resequencing buffer. To avoid timeouts at the TCP sender, our proposed solution
implements dynamic buffer sizes based on a \emph{buffering threshold} $T$,
sized as a function of the $\mathrm{MDI}$ and the bandwidths of the selected
paths for the flow.  If the number of packets buffered for a flow exceeds its
threshold, the buffer releases them all in order, ignoring gaps. This may
trigger some unnecessary retransmissions, but endpoints supporting SACK should
see only minimal impact.

Additionally, to protect against bursts of losses in the network, the next
expected sequence is further increased by a loss-recovery factor $\mathrm{LRF}$
after a threshold-triggered release. This causes the buffer to forward packets
with lower sequence numbers in their order of arrival, ignoring other lost
packets of the burst, until the new expected value is reached, thereby ignoring
any other missing packets from the loss burst. Experimental tests showed that a
value of 20 allowed the buffer to recover from bursty losses while limiting the
amount of out-of-order packets during this recovery period.

\subsection{Implementation Considerations}
\label{sec:implementation}

We implemented the WRR scheduler using the existing \emph{Select group} in Open
vSwitch. The resequencing buffer required the addition of a new group in Open
vSwitch, as well as a new OpenFlow message to configure it.  The code for these
modifications is available
online\footnote{\url{https://github.com/dariobanfi/ovs-multipath}}, as is that
of our Ryu-based
controller.\footnote{\url{https://github.com/dariobanfi/multipath-sdn-controller}}

While layer-2 path capacity was estimated using port statistics, a dynamic
layer-3 equivalent was not fully implemented---the controller currently needs
manual configuration of path capacities. We expect the switches could use
client traffic to implement methods such as packet
dispersion~\cite{dovrolis2004packet}. Such an approach is, however, beyond the
scope of this paper.

%% file: docs/evaluation.tex
\section{Performance Evaluation}
\label{sec:evaluation}

We evaluated MPSDN using both emulation and large-scale deployment on a
multihomed testbed. We first used emulation of an L2 topology to explore the
sensitivity of our approach to variations in conditions.  We then performed
use-case experiments in real-world L3 deployments to confirm the feasibility of
our solution. In both cases, we provide comparisons with MPTCP.

All measurements were done using Linux with default TCP parameters. In
particular this means that CUBIC was used as the congestion avoidance algorithm
for all TCP flows throughout this section.

\subsection{Emulation}

\begin{figure}[tb]
  \centering
  \includegraphics[width=\linewidth]{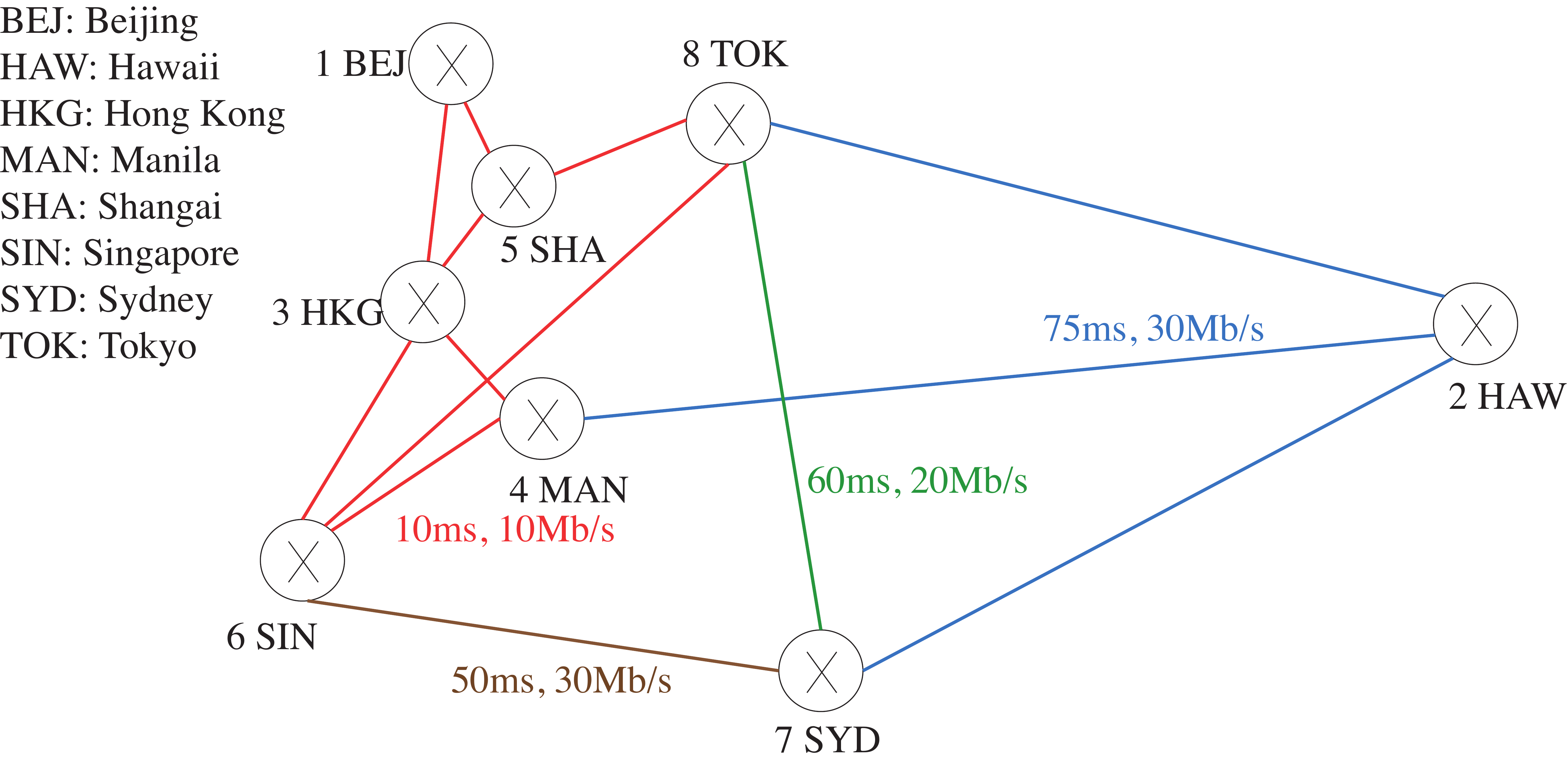}
  \caption{Topology model in our emulation (East Asia Internet Backbone).}
  \label{fig:evaluation_topology}
\end{figure}

We used Mininet~\cite{Lantz2010} to create an L2 topology mirroring the East
Asia Internet Backbone,\footnote{\url{http://maps.level3.com/default/}}, shown
in Figure~\ref{fig:evaluation_topology}.  As our setup could not emulate the
Gigabit speeds of the backbone, we scaled the capacities down. However, we chose
realistic delays between the routers, as estimated by probing their real
counterparts with ICMP echo requests.

\subsubsection{Throughput Measurements}

In the following experiments, we used \texttt{iperf}
3,\footnote{http://software.es.net/iperf/}
\texttt{netperfmeter}~\cite{dreibholznetperfmeter}, \texttt{netcat} and
\texttt{d-itg}~\cite{2004avallone_d-itg} to generate traffic. We measured flow
parameters (\texttt{cwnd}, \texttt{rtt}) with \texttt{ss} and \texttt{captcp}.

\begin{table}[bt]
  \centering
  \caption{Evaluation of throughput with MPSDN multipath forwarding.}
  \label{tab:paths_throughput}
  \begin{tabular}{ cccc }
    \toprule
    \textbf{Path}	& \textbf{Capacity}	& \textbf{Latency}	& \textbf{Throughput}		\\
    \midrule
    BEJ--SHA--TOK--HAW	& {\SImbps{10}}		& 95\,ms		& \multirow{2}{*}{\SImbps{18.2}}\\
    BEJ--HKG--MAN--HAW	& {\SImbps{10}}		& 95\,ms		\\
    \midrule
    TOK--SIN--SYD	& {\SImbps{10}}		& 60\,ms		& \multirow{2}{*}{\SImbps{26.8}}\\
    TOK--SYD		& {\SImbps{20}}		& 60\,ms		\\
    \bottomrule
  \end{tabular}
\end{table}

We measured the throughput our solution achieved without cross-traffic.
The results are shown in Table~\ref{tab:paths_throughput}. In this first 
scenario, the capacity was close to the aggregated bandwidth of the
single paths.

We also measured the throughput achieved with unbalanced latencies. The result
of these measurements is shown in \figurename~\ref{fig:mdi}. These measurements
were performed on a simple topology with just two direct paths.  One path has a
\SIms{25} latency, the other path's latency is increased to obtain the
different $\mathrm{MDI}$ values ($x$-axis). The measurements were done both
with and without enabling the resequencing buffer.  The effectiveness of the
buffer within a certain range of $\mathrm{MDI}$ values can be clearly observed.

\subsubsection{MDI cutoffs}

\begin{figure}[tb]
  \centering
  \includegraphics[width=\linewidth]{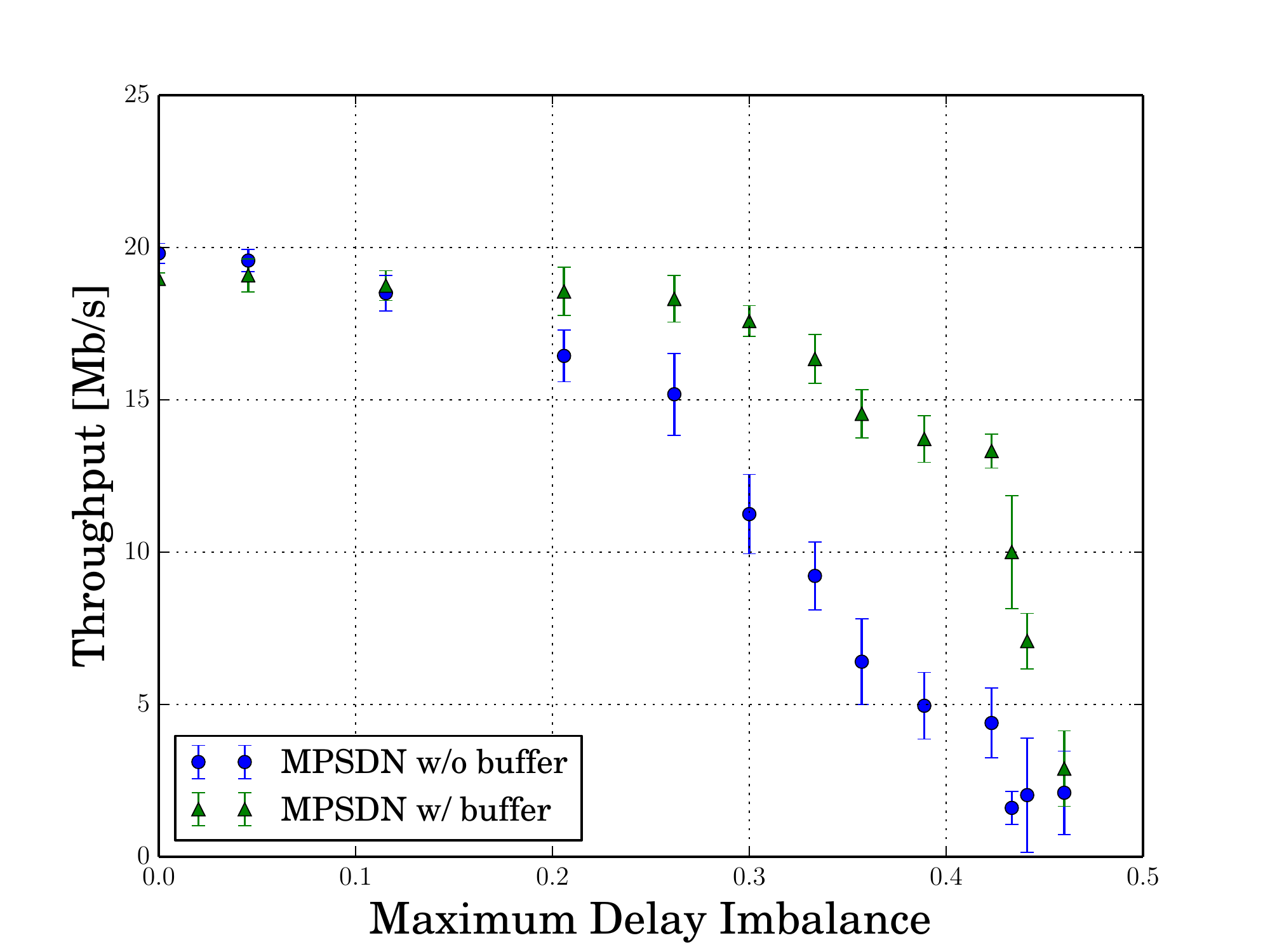}
    \caption{Impact of the maximum delay imbalance $\mathrm{MDI}$ on the aggregated
    throughput, with and without resequencing buffer (base latency \SIms{25},
    \SImbps{10} paths, 15-seconds \texttt{iperf})}
  \label{fig:mdi}
\end{figure}

\figurename~\ref{fig:mdi} also shows that using the resequencing buffer for
$\mathrm{MDI}$s beyond $0.15$ improves performance quite vastly, while for path
capacities beyond $0.4$ the aggregated bandwidth falls below the bandwidth of a
single path even when using the resequencing buffer.

\subsubsection{Intra-flow fairness}

We also verified that introducing MPSDN in a network does not have an adverse
effect on intra-flow fairness. We started 10 30-second \texttt{iperf}
transmissions over an MPSDN network and reported the flow throughput every
second. We computed Jain's fairness index~\cite{1984jain_fairness} for each
period. Overall, the mean fairness was 0.81 ($\sigma=0.042$), which we find to
be quite good.\footnote{The best fairness index would be 1, but anything above
  $0.5$ is considered ``reasonably fair''~\cite{rfc5348}.}

\subsubsection{Impact of path congestion on transport}

We then evaluated the resilience of our approach to dynamic congestion, both in terms
of path recomputation and transport resilience to changes.

\begin{figure}[tb]
    \centering
    \subfloat[Aggregated throughput]{
      \includegraphics[width=\linewidth]{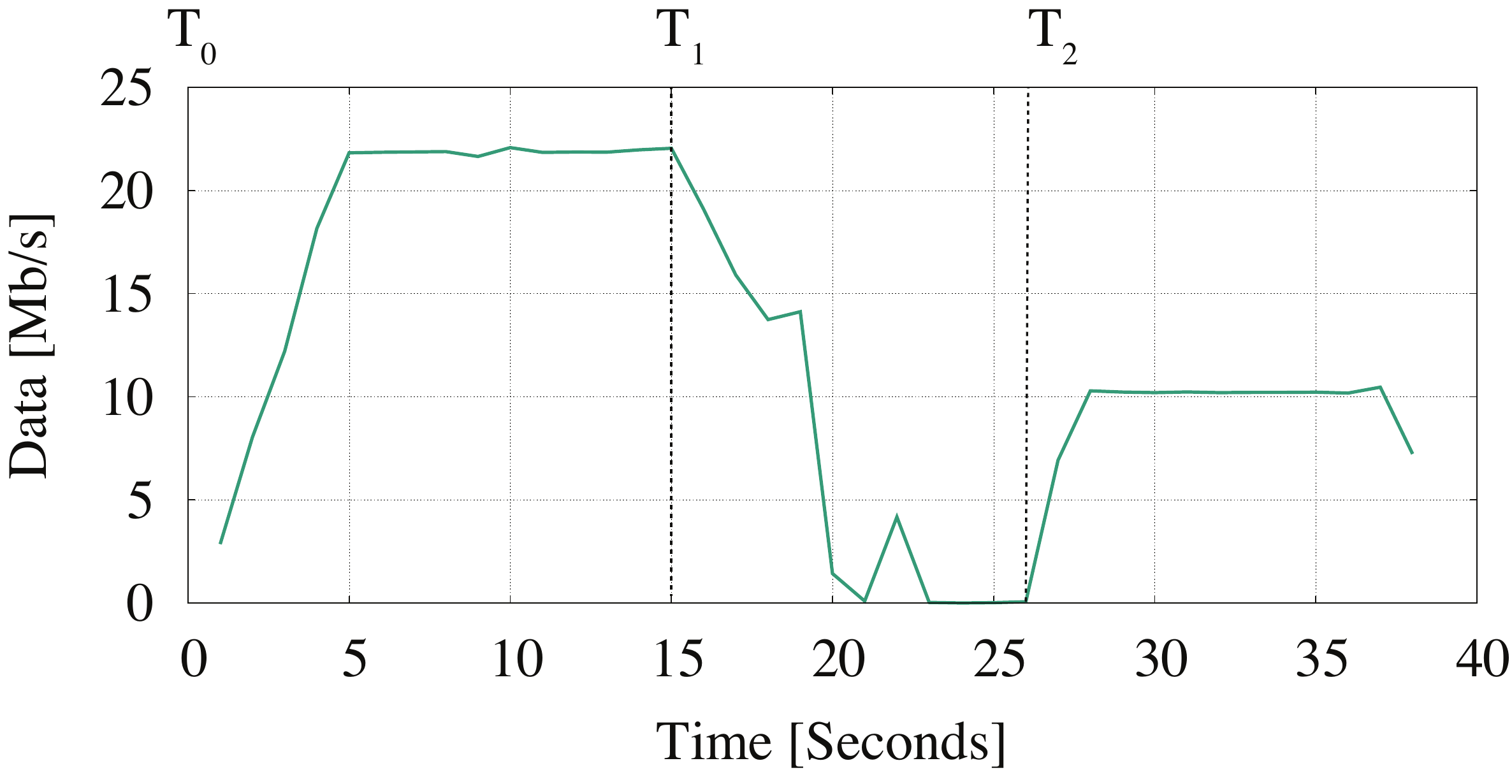}
      \label{fig:evaluation_congestion_tp}
    }

    \subfloat[Congestion window]{
      \includegraphics[width=\linewidth]{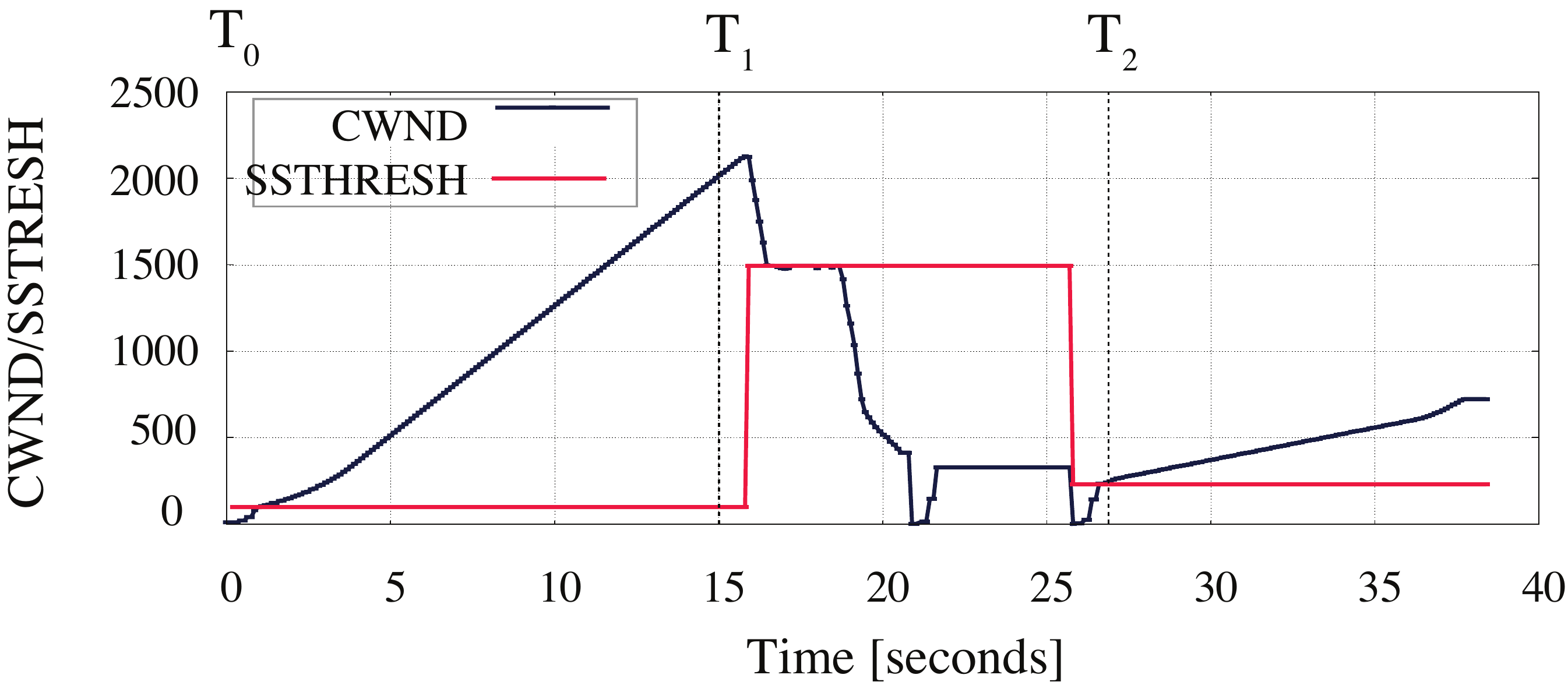}
      \label{fig:evaluation_congestion_cwnd}
    }
    \caption{Impact of transient congestion of one of the paths.}
\end{figure}

We used the same topology and source/destination as before. At time $T_1$, we
start an \texttt{iperf} UDP with a target rate of \SIgbps{1} to completely
saturate the link between Singapore (\emph{SIN}) and Sydney (\emph{SYD}).  At
$T_2$, our controller measures the available path capacity and recomputes 
bucket weights. As one path is completely congested, it switches to forwarding
on one path only.

Figure~\ref{fig:evaluation_congestion_tp} shows the impact of the path
congestion on the transport's throughput, where it quickly drops to $0$ before
our mechanism reconfigures the paths, after which the throughput slowly grows
to the new one-path capacity of 10\,Mbit/s.

While the controller adequately updated the path selection, the transport is
badly impacted during the congested period---even though only one path is
congested---and slow to respond after the path recomputation. This is due to
the TCP sender only maintaining a single congestion window for all the paths,
and reducing it drastically when losses start to occur on the congested path,
as shown in \figurename~\ref{fig:evaluation_congestion_cwnd}.

\subsubsection{Comparison with MPTCP}

MPTCP and MPSDN differ in the requirements they impose on implementing systems:
multi-homing in the case of MPTCP, and SDN support with measurement capability
for MPSDN. Nonetheless, they share the same objective of capacity aggregation.
We therefore compared the goodput achieved by our solution to MPTCP's, in
systematic experiments varying the delay on the second path.

We set up a basic topology composed of just two hosts. For MPTCP, the hosts are
multi-addressed. For MPSDN, each host has only one IP address, but there are
two available paths in the network. For MPTCP, we use two subflows and the
default scheduler. The sender starts a 30 seconds transmission; the
application-layer goodput is measured at the receiver.

\begin{figure*}[tb]
  \centering

  \subfloat[$c_{p_2}=10\text{\,Mbit/s}\quad d_{p_2}=25\text{\,ms}$]{\includegraphics[width=.33\linewidth]{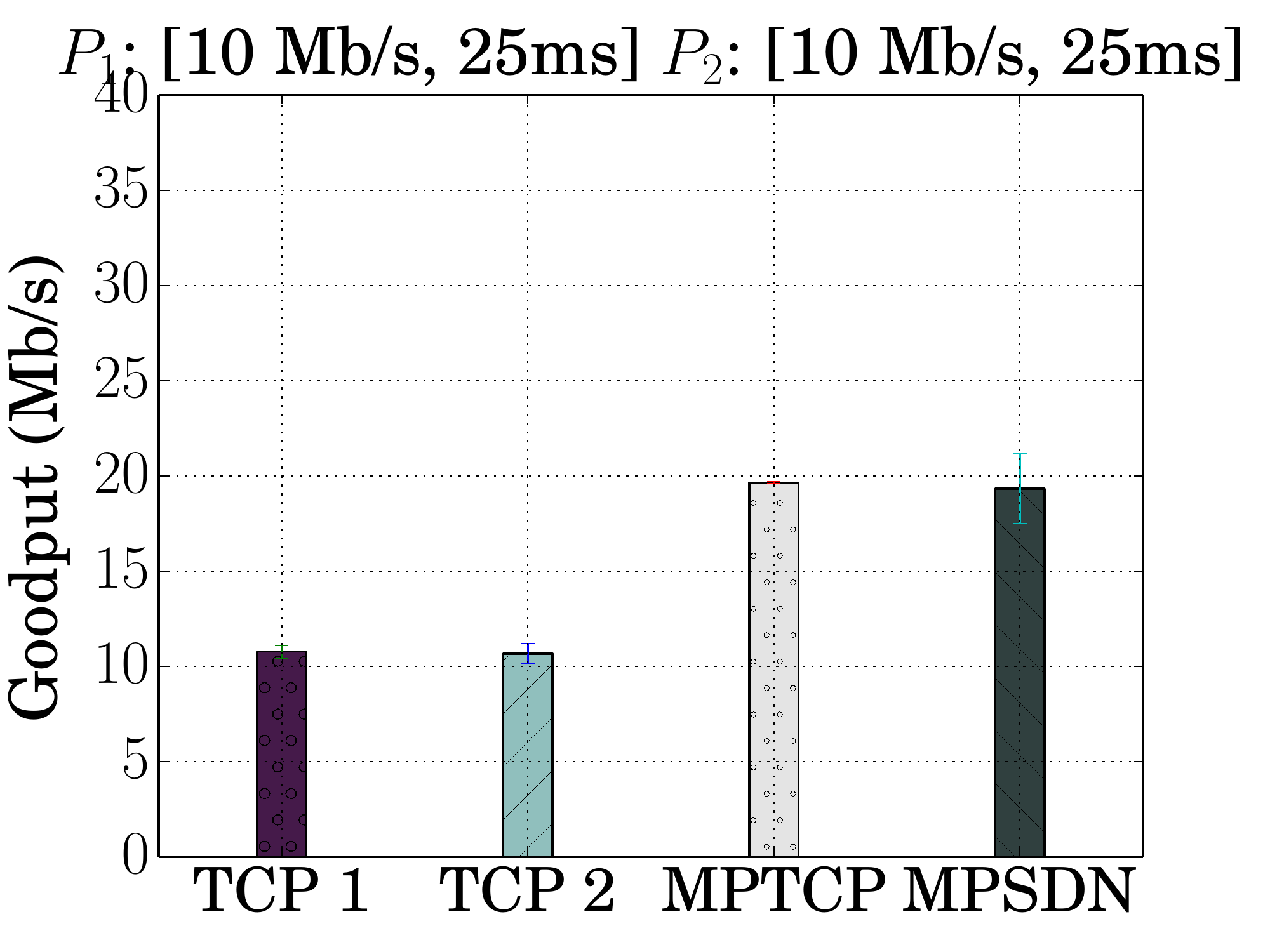}}
  \subfloat[$c_{p_2}=10\text{\,Mbit/s}\quad d_{p_2}=50\text{\,ms}$]{\includegraphics[width=.33\linewidth]{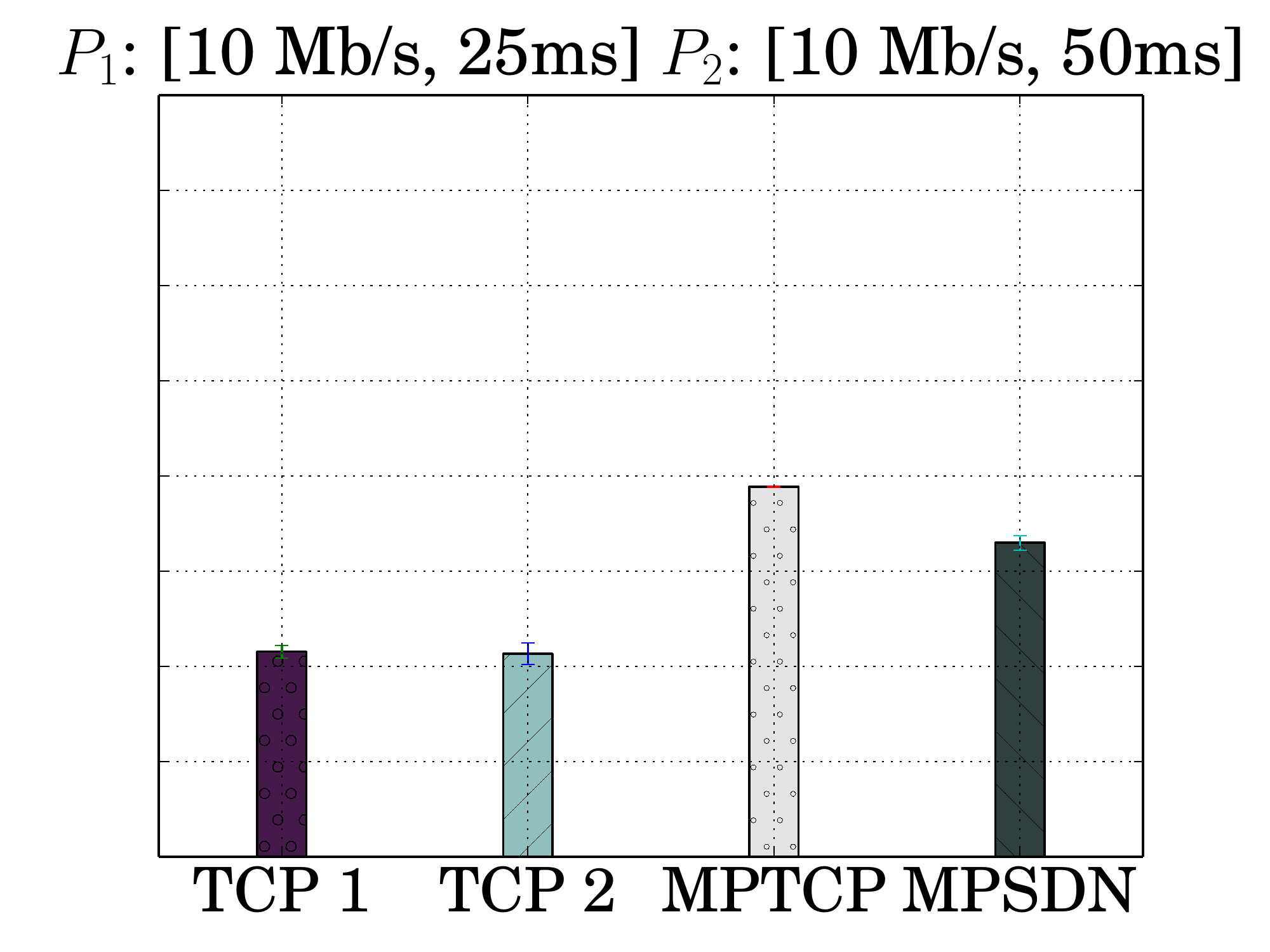}}
  \subfloat[$c_{p_2}=10\text{\,Mbit/s}\quad d_{p_2}=100\text{\,ms}$]{\includegraphics[width=.33\linewidth]{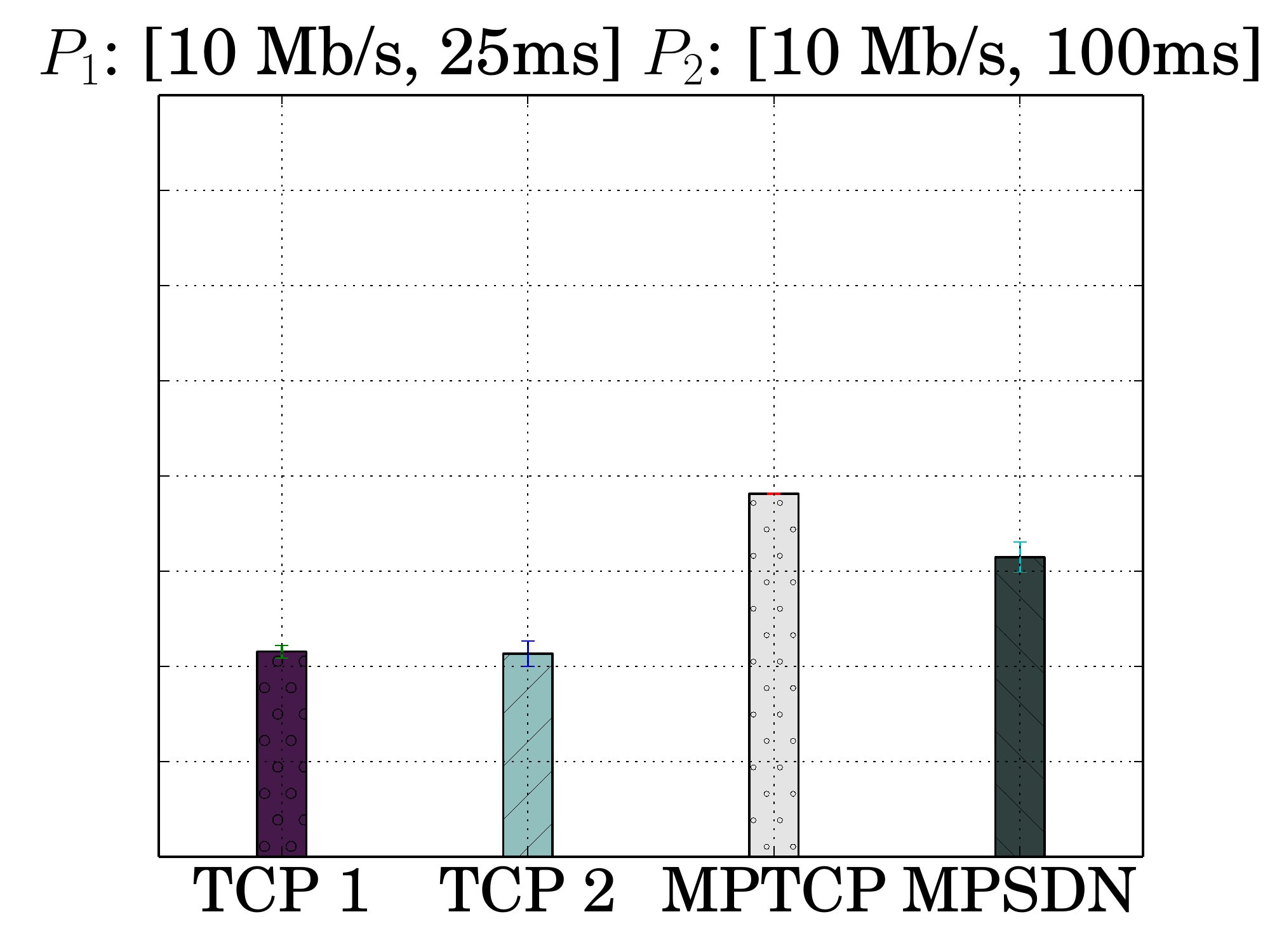}}

  \caption{Comparison of MPSDN goodput to MPTCP with varying path delays. Error
  bars show the standard deviation of 10 runs.}

  \label{fig:mpsdn_vs_mptcp}
\end{figure*}

Figure~\ref{fig:mpsdn_vs_mptcp} shows the TCP goodput for the single paths and
compares it to MPTCP and MPSDN performance.  In  sub-figures (b) and (c), which
have a high delay difference (\SIms{25} and \SIms{50} corresponding to an
$\mathrm{MDI}$ of $0.17$; \SIms{25} and \SIms{100} corresponding to an
$\mathrm{MDI}$ of $0.3$) the reordering buffer is used.

Our results show that MPSDN performance remains close to that of MPTCP when the
paths are balanced (although with a higher variance) and performs slightly
worse, but still comparable, when the delay differences are high.

\subsection{Real-world deployment}

We now verify that our proposal is usable in real world deployments. The most
notable difference is that, instead of an L2 topology, we now consider an L3
network where we only control the edge switches. Apart from quantitative
measurements, our goal is also to qualitatively explore the deployability of our
solution over the real Internet.

We deployed our MPSDN solution on the
NorNet\footnote{\url{https://www.nntb.no/}} Core testbed, which offers
distributed, multihomed, and programmable nodes~\cite{Dreibholz}, where static
IP tunnels are established to form a full mesh between nodes, and packets are
routed based on their source/destination address.

We ran our experiments on Ubuntu 14.04 LTS virtual machines with kernel
3.13.0-68-generic, 1\,GB RAM, and 2.60\,GHz CPUs. The VMs were multi-addressed
and used the aforementioned IP tunnelling. We simply installed our modified
Open vSwitch directly on the VMs and used it to route the traffic.  This allows
the application to create normal TCP connections and keeps the multipath
splitting transparent.

We chose the sites at the Simula Research Laboratory near Oslo (NorNet's home)
and one in Longyearbyen, just 1300\,km from the North pole, in the Svalbard
archipelago. 

The switch on the sender side was configured to rewrite the layer-3
source/destination addresses to trigger the Weighted Round Robin Scheduling and
forward packets onto their selected path. The receiving switch performed the
reverse address mapping.

\subsubsection{TCP Goodput}

We first tested the scheduling without any reordering buffer between two
endpoints with paths of equal capacity to determine how many packets would
arrive out-of-order and cause performance degradation.  We used two of the
multiple paths/ISP combinations between both endpoints, which had at least
10\,Mbit/s of capacity. As discussed in Section~\ref{sec:implementation}, we
manually set the weights for both paths in the scheduler. We set them to equal
values.  Both paths have RTTs around 40\,ms.

\begin{figure}[tb]
  \centering
  \includegraphics[width=\linewidth]{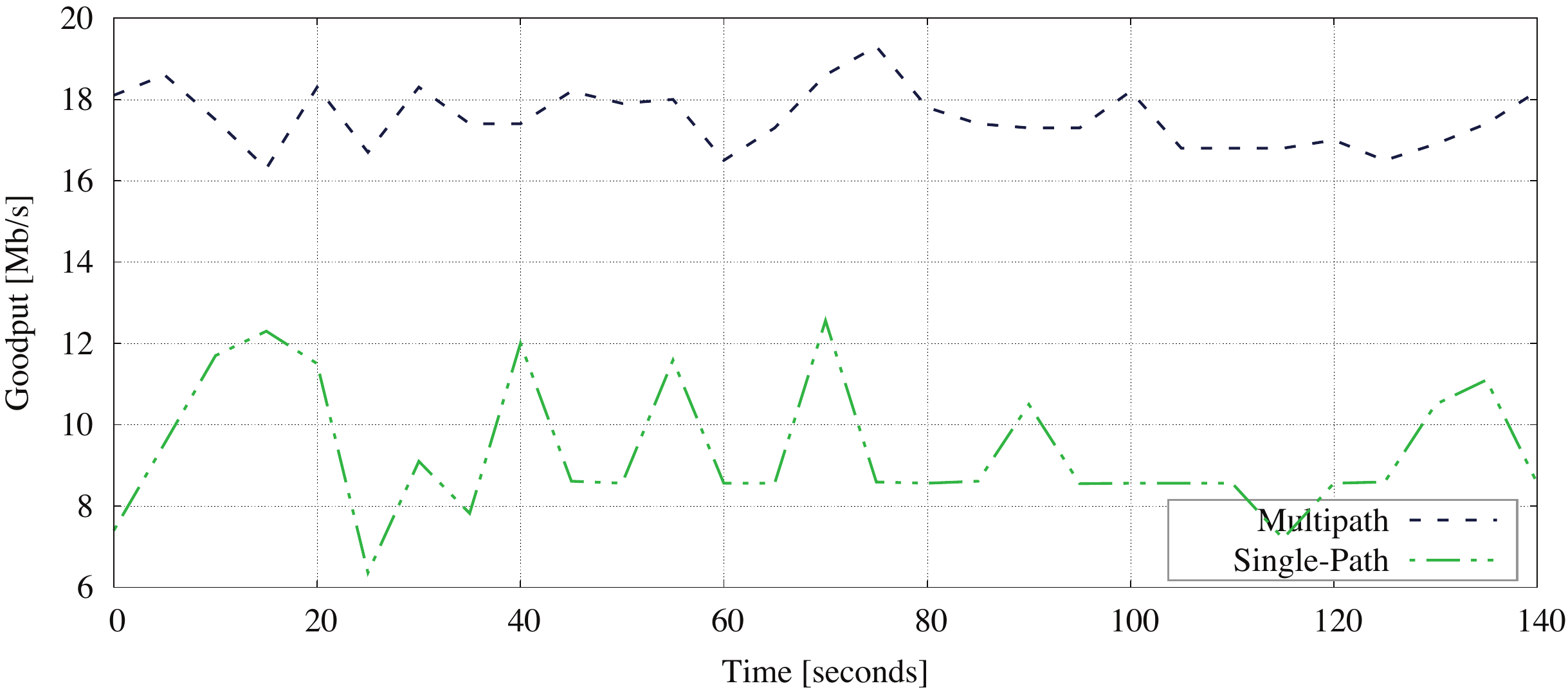}
  \caption{\texttt{iperf} throughput: single-path vs. multipath.}
  \label{fig:iperf3_nornet}
\end{figure}

Figure \ref{fig:iperf3_nornet} shows the goodput over 140 seconds as measured
by \texttt{iperf} 3 periodic reports, with the default settings.  Multipath
forwarding succeeds in aggregating paths capacities, resulting in a roughly
doubled throughput, compared to single-path.

\begin{figure}[tb]
  \centering
  \includegraphics[width=\linewidth]{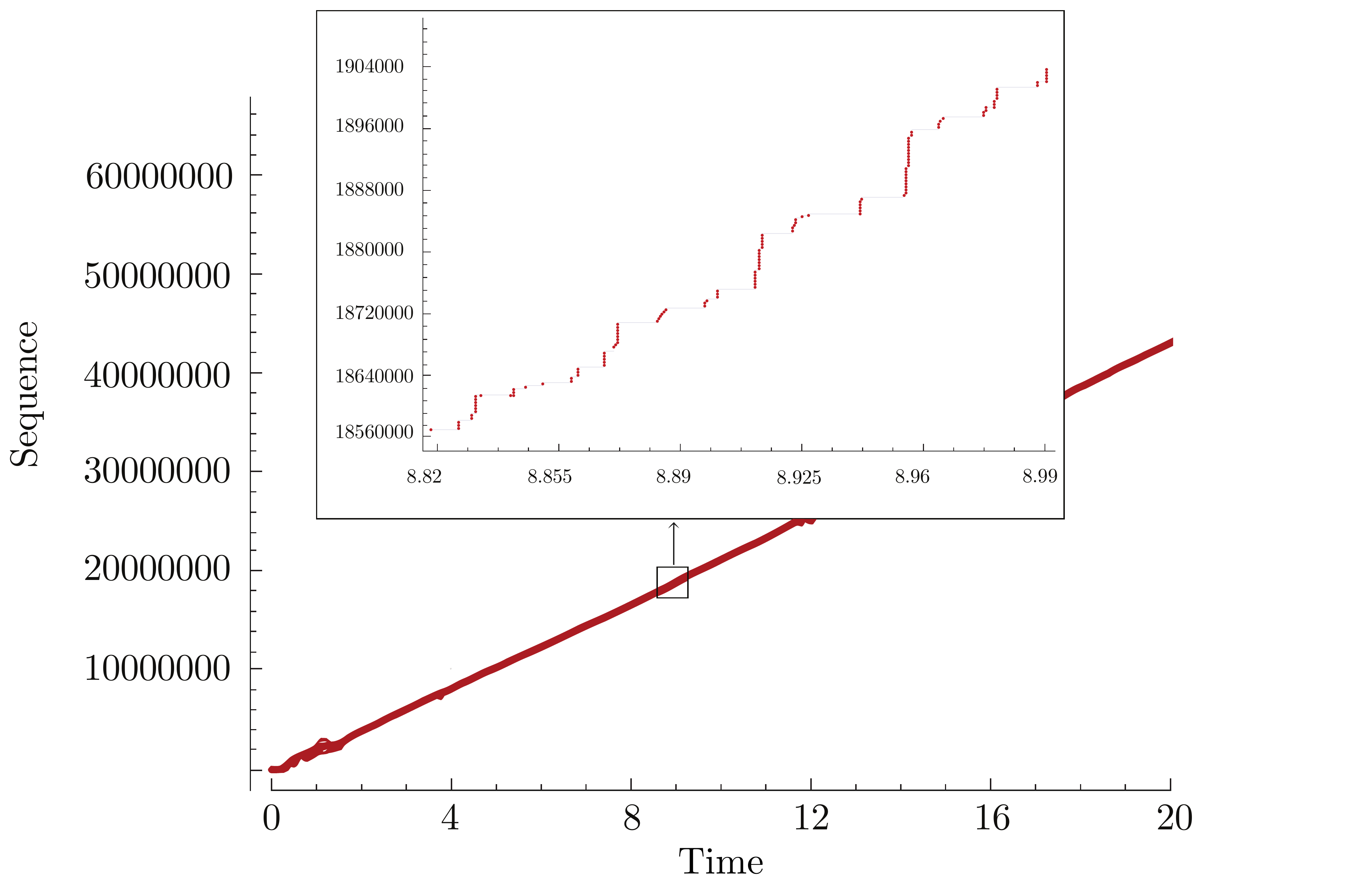}
  \caption{TCP sequence numbers at the receiver.}
  \label{fig:sequences_nornet}
\end{figure}

As Figure~\ref{fig:sequences_nornet} shows, the TCP sequence numbers at the
receiving end are growing almost monotonically, showing only very light packet
reordering. It is interesting to note the jagged profile of the curve, where
bursts of packets arrive at different rates, depending on which paths they had
been forwarded on.

\subsubsection{Application use-cases}

We continued with the same configuration, and experimented with a number of
different application workloads.  We first tested with the same setup, but also
launched \texttt{netperfmeter} using TCP and the default flow settings, which
attempts to maximise the throughput. We then ran two other experiments. We used
Python's
\texttt{SimpleHTTPServer}\footnote{\url{https://docs.python.org/2/library/simplehttpserver.html}}
module and \texttt{wget}\footnote{\url{https://www.gnu.org/software/wget/}} to
simulate the transfer of a \SIMB{70} file over HTTP, and over FTP using
\texttt{vsftpd}.\footnote{\url{https://security.appspot.com/vsftpd.html}}

\begin{figure}[tb]
  \centering
  \includegraphics[width=\linewidth]{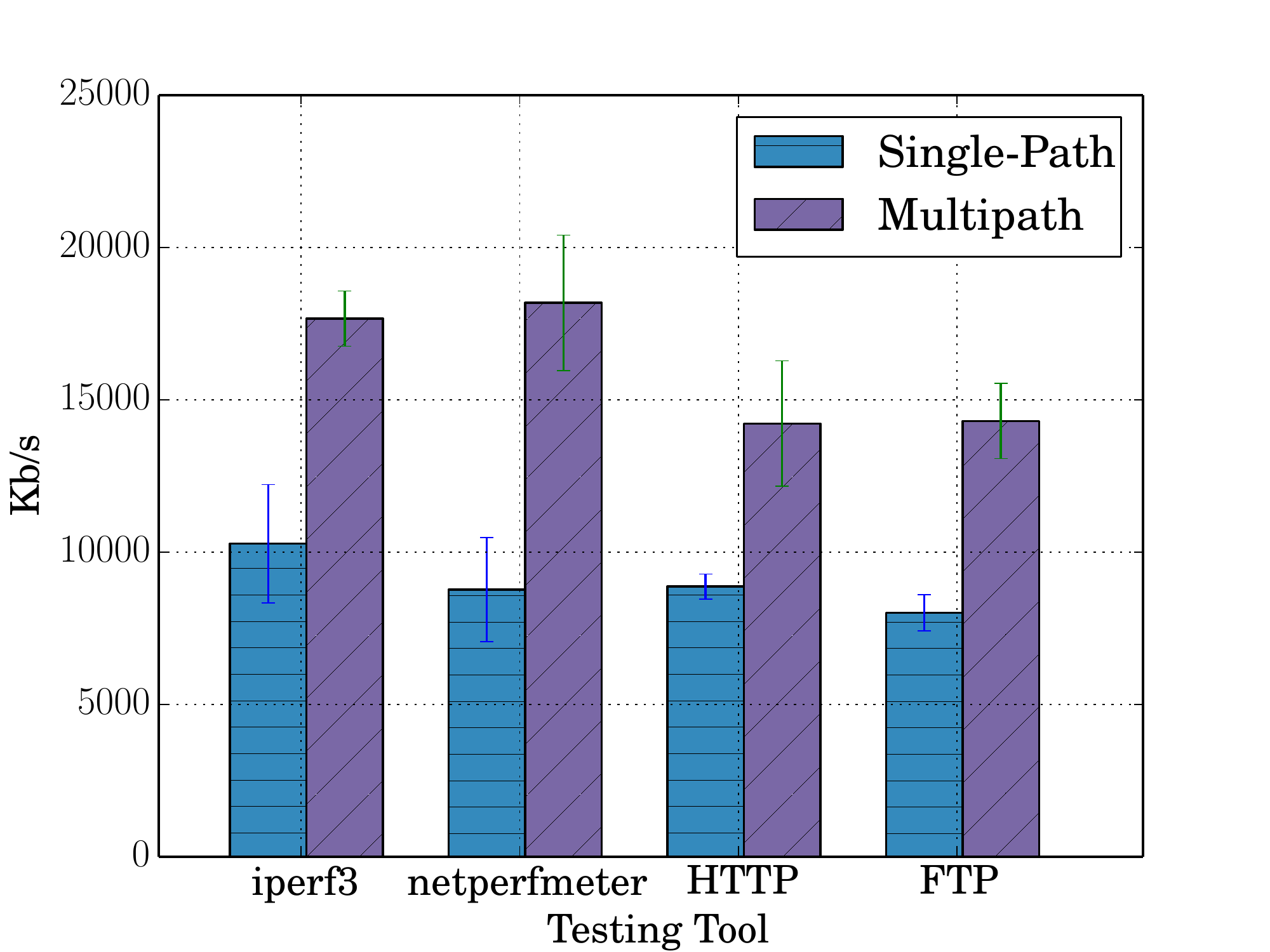}
  \caption{Multipath throughput with different traffic types.}
  \label{fig:traffic_nornet}
\end{figure}

Figure~\ref{fig:traffic_nornet} shows our results. Multipath forwarding
consistently delivered on its promise of capacity aggregation. This is best
shown for applications which actively attempt to saturate the network capacity,
but all significantly benefit from MPSDN.

%% file: docs/discussion.tex
\section{Discussion and Lessons Learnt}
\label{sec:discussion}

In this section, we reflect on the proposed architecture, influence of the
design choices, and resulting performance and usability of MPSDN.

\subsubsection{Impact of buffer and MDI} 

The $\mathrm{MDI}$ proved to perform quite well in our experiments. However,
all our measurements have a somewhat similar latency (with the faster path
usually having a latency of either \SIms{10} or \SIms{25}). We later realised
that using a purely relative measurement for the imbalance does not allow the
$\mathrm{MDI}$ to work equally well for all latencies. It needs to be refined
to also incorporate the absolute difference between the latencies.

\subsubsection{Path selection} In the evaluated version of our proposal, we did
not consider a dynamic and continuous estimation of the paths' capacity, but
rather focused on the feasibility of our solution in a stable environment. We
showed in this context that, in conjunction with the $\mathrm{MDI}$ metric, it
was possible to identify suitable complementary paths. Nonetheless, as future
work, we plan to investigate the possible addition of a measurement mechanism
(active or passive) to estimate these capacities. In particular, we aim to
determine the trade-off between adding more measurements and accidentally
contributing to congestion. It might also be worthwhile to replace a
currently-congested path with another, uncongested, path.

\subsubsection{Impact on Vanilla TCP}

Our solution successfully enables transparent capacity aggregation by
scheduling bursts of packets on different paths, and prevents spurious
retransmissions by reordering datagrams before delivering them to the end node.
Nonetheless, TCP's control loop can become disrupted due to transient issues on
any single path, leading to performance degradation for the whole transfer.

This is due to the fact that the TCP has no knowledge of the use of multiple
paths. Its RTT estimate is that of the longer path and reordering buffer, while
its congestion window covers the aggregated capacity. In case one path
experiences a spike in delays, or a burst of losses, TCP will react by reducing
its sending rate \emph{for the whole transfer}. As a result, only paths of
similar characteristics (as determined by metrics such as the $\mathrm{MDI}$)
will aggregate well, but the throughput will be very sensitive to the
performance of the worst path.

\subsubsection{Comparison to MPTCP}

Even without a reordering buffer, our in-network multipath solution achieves a
very good aggregated bandwidth and similar goodputs as MPTCP while not
requiring end-host support.

An SDN solution, with its advantages of being network stack-agnostic, can
achieve a performance that is similar to that of MPTCP. While MPTCP's challenge
is endpoint support, the challenge with MPSDN lies in determining the
parameters for path setup and packet reordering.

\subsubsection{Ease of deployability}

Our MPSDN proposal reduces the deployability issues seen with MPTCP.  While
each end-host needs to be separately enabled to support MPTCP, MPSDN only
requires leaf networks to deploy at least one edge switch supporting our
extensions to provide multipath connectivity from all hosts on that network to
any other MPSDN-enabled network. Some deployment considerations were however
not addressed, such as when two MPSDN networks are not under the jurisdiction
of the same controller. Access control and delegation in SDN is beyond the
scope of this paper, but can be adequately addressed by a broader research
agenda in SDN~\cite[\eg,][]{2014baldin_resource_delegation_sdn}.

%% file: docs/conclusion.tex
\section{Conclusion}
\label{sec:conclusion}

We have presented a solution to enable the use of multiple paths in a layer-2
or -3 topology. The main objective is to use alternate paths in parallel to
aggregate capacity and provide higher goodputs.  Unlike solutions such as MPTCP
or CMT-SCTP, our approach leverages an SDN infrastructure to provide path
selection, packet scheduling, and packet reordering in the network, without the
need to modify the endpoints. We have evaluated the solution in a range of
emulated scenarios and showed that it is able to adequately provide
capacity-aggregation benefits that are similar to what MPTCP achieves.  We have
also demonstrated the deployability of the solution in a real multi-homed
scenario over the Internet.

Our work highlighted the need that the various aspects of multipath transfer
are addressed in the right layer---path discovery and selection belongs in the
network, but the transport needs to be aware of the existence of multiple paths
and manage them separately---and a richer communication between those layers to
support it. Future work should study how this split can best be achieved.
Unfortunately, TCP/IP networks are poorly equipped for a lightweight upgrade
that could unlock the full potential of multiple paths.